\documentclass[conference, 10pt]{IEEEtran}
\IEEEoverridecommandlockouts
\usepackage{graphicx}
\usepackage{multirow}
\usepackage{epstopdf}
\usepackage{amsmath}
\usepackage{amssymb}
\usepackage[noend]{algpseudocode}
\usepackage[linesnumbered,ruled,vlined]{algorithm2e}
\usepackage{amsfonts}
\usepackage{hyperref}
\usepackage{subfig}
\usepackage{listings}
\usepackage{xcolor}
\usepackage{xspace}
\usepackage{pdfpages}
\usepackage{framed}
\usepackage{mathtools}
\usepackage[most]{tcolorbox}
\usepackage{floatrow}
\usepackage{epstopdf}
\usepackage{multirow}
\usepackage{wrapfig}
\usepackage[shortlabels]{enumitem}
\setlist{nosep,noitemsep,after=\vspace{5pt}}
\newfloatcommand{capbtabbox}{table}[][\FBwidth]
\newcommand{\toolname}{\textsc{Hydra}{}}
\newcommand{\corral}{\textsc{Corral}\xspace}
\newcommand{\todo}[1]{\textcolor{red}{TODO: #1}}
\newcommand{\figref}[1]{Figure~\ref{Fi:#1}}
\newcommand{\sectref}[1]{Section~\ref{Se:#1}}

\newcommand{\algoref}[1]{Algorithm~\ref{Alg:#1}}
\newcommand{\eqnref}[1]{Equation~\ref{Eqn:#1}}
\newcommand{\lineref}[1]{Line~\ref{Ln:#1}}
\newcommand{\linesref}[2]{Lines~\ref{Ln:#1} to~\ref{Ln:#2}}
\newcommand{\blocks}{\textit{blocks}}
\newcommand{\bsucc}{\textit{succ}}
\newcommand{\bpred}{\textit{pred}}
\newcommand{\callsites}{\textit{callsites}}
\newcommand{\calltree}{\textit{calltree}}
\newcommand{\decisions}{\textit{decisions}}
\newcommand{\cvar}{\textit{control-variable}}
\newcommand{\interfacevar}{\textit{interface-variables}}
\newcommand{\pVC}{\text{pVC}}
\newcommand{\mr}{\textit{must-reach}}
\newcommand{\ma}{\textit{must-avoid}}

\newcommand{\solver}{\mathcal{S}}
\newcommand{\Omit}[1]{}
\newcommand{\callServer}{SendSync}
\newcommand{\callServerA}{SendAsync}

\usepackage[nodisplayskipstretch]{setspace}

\lstdefinelanguage{boogie}{%
  basicstyle=\tt,
  classoffset=0,
  keywords={var,const,reg,proc,procedure,skip,assume,assert,goto,havoc,if,then,else,%
    while,do,call,return,returns,post,await,ewait,yield,
    let,in,and,or,true,false,
    for,from,to,
    future, touch,
    fork, rfork, join,
    async, finish,
    spawn, sync, inlet,
    type,
    eventloop,
    foreach,
    int,bool
  },
  keywordstyle=\bf,
  classoffset=1,
  morekeywords={g,l},
  keywordstyle=\tt,
  classoffset=0,
  basicstyle=\tt,
  commentstyle=\itshape,
  morecomment=[l]{//},
  morecomment=[s]{/*}{*/},
  mathescape=true,
  escapeinside=`'
}

\lstnewenvironment{boogie}{%
  \lstset{%
    language={boogie},
    columns=flexible,
    breaklines=true,
    tabsize=1,
  }
}{}

\SetAlFnt{\small}
\algnewcommand{\IIf}[1]{\State\algorithmicif\ #1\ \algorithmicthen}
\algnewcommand{\EndIIf}{\unskip\ \algorithmicend\ \algorithmicif}

\begin{document}

\title{Distributed Bounded Model Checking\\}

\author{
\IEEEauthorblockN{Prantik Chatterjee\IEEEauthorrefmark{1},
Subhajit Roy\IEEEauthorrefmark{1},
Bui Phi Diep\IEEEauthorrefmark{2},
Akash Lal\IEEEauthorrefmark{3}
}
\IEEEauthorblockA{\IEEEauthorrefmark{1}IIT Kanpur,
\{prantik, subhajit\}@cse.iitk.ac.in
}
\IEEEauthorblockA{\IEEEauthorrefmark{2}Uppsala University,
bui.phi-diep@it.uu.se
}
\IEEEauthorblockA{\IEEEauthorrefmark{3}Microsoft Research,
akashl@microsoft.com
}
}

\maketitle

\begin{abstract}
Program verification is a resource-hungry task. This paper looks at the problem of parallelizing 
SMT-based automated program verification, specifically bounded model-checking, so that it can be 
distributed and executed on a cluster of machines. We present an algorithm that
dynamically unfolds the call graph of the program and frequently splits it to create sub-tasks
that can be solved in parallel. The algorithm is adaptive, controlling the splitting rate
according to available resources, and also leverages information from the SMT solver to split where
most complexity lies in the search. We implemented our algorithm by modifying \corral, the verifier used by
Microsoft's Static Driver Verifier (SDV), and evaluate it on a series of hard SDV benchmarks.
\end{abstract}

\section{Introduction}
\label{Se:Introduction}

Program verification has a long history of over five decades and it has been
consistently challenged over this entire duration by the continued increase in the
size and complexity of software. As the efficiency of techniques and solvers has
increased, so has the
amount of software that is written. For this reason, \textit{scalability}
remains central to the applicability of program verification in practice.

This paper studies the problem of automated program verification. In particular,
we consider Bounded Model Checking (BMC) \cite{DBLP:conf/dac/ClarkeKY03}: 
the problem of reasoning over the entire space of program inputs but only over a subset
of program paths, typically up to a bound on the number of loop iterations and
recursive calls. BMC side-steps the need for (expensive and undecidable) 
inductive invariant generation and instead directly harnesses the power of
SAT/SMT solvers in a decidable
fragment of logic. BMC techniques are popular; they are implemented in most
program verification tools today \cite[Table~5]{DBLP:conf/tacas/Beyer19}.

Our goal is to scale BMC by parallelizing the verification
task and distributing it across multiple machines to make use of larger compute and
memory resources. The presence of several public cloud providers  has made it easy to
set up and manage a cluster of machines. While this distributed platform is
available to us, there is a shortage of verification tools that can exploit it.

\noindent \textit{Parallelizing BMC.~}
BMC works by generating logical encodings, often called \textit{verification
conditions} or VCs, for a subset of program paths that are then fed to an SMT
solver to look for potential assertion violations in the program. We aim to retain the same 
architecture, where we continue to use the SMT solver as a black-box, but 
generate multiple different VCs in parallel to search over disjoint sets of 
program paths. This allows us to directly consume future improvements in SMT
solvers, retaining one of the key advantages of BMC.

Our technique works by \textit{splitting} the set of program paths into disjoint subsets that are then
searched independently in parallel. The splitting is done by simply picking a control node
and considering $(a)$ the set of paths that go through the node, and
$(b)$ the set of paths that do not. Splitting can happen multiple times. 
The decisions of \textit{what} node to split and \textit{when} to split are both 
taken dynamically by our technique. We refer to the
BMC problem restricted to a set of splitting decisions (i.e., nodes that must be
taken, and nodes that must be avoided) as a \textit{verification partition}.

Verification starts by creating multiple processes, each of which have access to 
input program and are connected over the network. One process is designated as
the \textit{server} while the rest are called \textit{clients}. 
The search starts sequentially on one of the clients that applies
standard BMC on the input program. At some point in time, which is controlled by
the \textit{splitting rate}, the client chooses a splitting node, thus creating
two partitions. The client continues verification on one of the partitions, and
sends the other partition to the server. The server is only responsible for coordination; it does not
do verification itself. It accumulates the partitions (represented as a set of splitting
decisions) coming in from the clients and farms them off to idle clients for
verification. Clients can split multiple times. This continues until a client reports a counterexample (in
which case, it must be a counterexample in the original program) or the server
runs out of partitions and all clients become idle (in which case, the BMC
problem is concluded as \textit{safe}). 

The splitting rate is adjusted according to the current number of idle client:
it is reduced when all clients are busy, and then increased as more clients
becomes available. 

Splitting has some challenges that we illustrate using the 
following snippet of code.
\begin{lstlisting}[language=boogie,basicstyle=\scriptsize]
procedure main() {
   var x := 0;
   if (...) { call foo(); x := 1; }
   if (...) { call bar(); }
   if (...) { call baz(); }
   assert(x == 1 || expr);
}
\end{lstlisting}
Suppose that the assertion at the end of \texttt{main} is the one that we
wish to verify (or find a counterexample) and all uses of variable \texttt{x}
are shown in the snippet. The \texttt{main} procedure calls multiple other
procedures, each of which can manipulate global variables of the program (not shown).
In this case, if we split on the call to \texttt{foo}, 
then one partition (the one that must take \texttt{foo}) becomes trivial: it 
is easy to see that the assertion holds in that partition, irrespective of what happens in the
rest of the program. We refer to this as a \textit{trivial} split. Each split
incurs an overhead when a partition is shipped to another client where the
verification context for that partition must be set up from scratch. Trivial splits are troublesome
because they accumulate this overhead without any real benefits in trimming down
the search. Unfortunately, it is hard to avoid trivial splits altogether because it can
involve custom (solver specific) reasoning (e.g., the fact that
variable \texttt{x} is not modified outside of \texttt{main}). Our technique
instead aims to reduce the overhead with splitting when possible. The server
prioritizes sending a partition back to the client that generated it. Each
client uses the incremental solving APIs of SMT solvers to remember backtracking 
points of previous splits that it had produced. This allows a client to get
setup for one of its previous partitions much faster, thus reducing overhead.

Next, consider splitting on the call to \texttt{bar}. In this case, both of the
generated partitions must still reason about \texttt{baz} because taking or avoiding
\texttt{bar} has no implications on the call to \texttt{baz}. If \texttt{bar}
turns out to be simple, while most of the complexity lies inside \texttt{baz}, 
then both partitions will end up doing the same work and diminish the benefits
of parallelization. In this case, we rely on extracting information from the
solver (via an unsat core) to make informed splitting choices and avoid duplicating
work across partitions.

\noindent \textit{Implementation.~} 
We have implemented our technique in a tool called $\toolname$. The sequential
BMC technique used by $\toolname$ is \textit{stratified
inlining} (SI) \cite{corral}, also referred to as \textit{inertial refinement}
\cite{DBLP:conf/fmcad/Sinha10}. SI incrementally builds the VC of a program by
lazily inlining procedure calls. $\toolname$ keeps track of the expanding
VC, and frequently splits it by picking a splitting node that has already been
inlined in the VC. 

We evaluated $\toolname$ on Windows Device Driver benchmarks obtained using
the Static Driver Verifier \cite{SDV,sdvurl}. These benchmarks extensively exercise the
various features of C such as heaps, pointers, arrays, bit-vector operations,
etc. \cite{DBLP:conf/sigsoft/LalQ14} and take more than $11$ days to verify in a sequential
setting.

The contributions of this paper are as follows:

\begin{itemize}
    \item We propose a distributed design to enable solving
	    large verification problems on a cluster of machines (\sectref{Client} and
	    \sectref{Server});
    \item We design a \textit{proof-guided} splitting strategy that
	    enables a lazy, semantic division of the verification task (\sectref{SplittingHeuristic}
	    and \sectref{SplitRate});
    \item We implemented our design in a tool called {\toolname} that achieves a 20$\times$
      speedup on 32 clients, solving $30\%$ additional benchmarks on which the
      sequential version timed out (\sectref{Experiments}).
\end{itemize}

The rest of the paper is organized as follows. \sectref{Background} covers background on VC
generation and the stratified inlining algorithm. \sectref{Algorithm} discusses on how the search is decomposed for parallel exploration while \sectref{DesignImplementation} presents the design of
$\toolname$. \sectref{Experiments} presents an evaluation of
$\toolname$ and \sectref{RelatedWork} discusses related work. 


\section{Background}
\label{Se:Background}

We describe our techniques on a
class of \textit{passified} imperative programs. Such a program can have multiple
procedures. Each procedure has a set of labelled basic blocks, where each block contains
a list of statements followed by a \textbf{goto} or
a \textbf{return}. A statement can only be an \textbf{assume} or a procedure
\textbf{call}. A procedure can have any number of formal input arguments and local
variables. Local variables are assumed to
be non-deterministically initialized, i.e., their initial value is unconstrained.
An \textbf{assume} statement takes an arbitrary expression over the variables in
scope. An example program is shown in \figref{ExampleProgram}. A \textbf{goto} statement takes multiple
block labels and non-deterministically jumps to one of them.

Passified programs do not have global variables, return parameters of
procedures, or assignments. These restrictions are without loss of generality
because programs with these features can be easily converted
to a passified program \cite{DBLP:conf/rp/LalQ13};
such conversion is readily available in tools like \textsc{Boogie}~\cite{boogie}.
We also leave the expression syntax unspecified: we only require that
expressions can be directly encoded in SMT.
Our implementation uses linear arithmetic, fixed-size bit-vectors,
uninterpreted functions, and extensional arrays. This combination is sufficient
to support C programs \cite{DBLP:conf/sigsoft/LalQ14,conf/popl/LahiriQ08}.

We aim to solve the following safety verification problem: given a
passified program $P$, is the end of its \texttt{main} reachable, i.e., is there an
execution of \texttt{main} that reaches its \textbf{return} statement?
This question is answered \textsc{Yes} (or \textsc{UnSafe}) by producing such an
execution and the answer is \textsc{No} (or
\textsc{Safe}) if there is no such execution. Furthermore, we only consider a
\textit{bounded} version of problem where $P$ cannot have loops or recursion.
(In other words, loops and recursive calls must be unrolled up to a
fixed depth.) This problem is decidable with NEXPTIME
complexity \cite{DBLP:conf/rp/LalQ13}.
We next outline VC generation for single-procedure (\sectref{SingleProcVC}) and
multi-procedure (\sectref{StratifiedInlining}) programs.


\begin{figure}[t]
    \begin{minipage}{0.42\textwidth}
  {
\begin{lstlisting}[language=boogie,basicstyle=\scriptsize]
procedure main() {
  int x, y, z; bool c;
  L0: goto L1, L2;
  L1: assume c;
      call foo(x,z);
      goto L3;
  L2: assume !c;
      call bar(x,z);
      goto L3;
  L3: call baz(y);
      goto L4;        
  L4: assume z != 0
      return;
}
procedure baz(int y) {
  L10: assume y == 3;
       return;
}

\end{lstlisting}}
    \end{minipage}~~
    \begin{minipage}{0.45\textwidth}
  {
\begin{lstlisting}[language=boogie,basicstyle=\scriptsize]
procedure foo(int x, int z) {
  bool d;
  L5: goto L6, L7;
  L6: assume d;
      assume z == x + 1;
      goto L8;  
  L7: assume !d;
      assume z == x - 1;
      goto L8;
  L8: return; 
}

procedure bar(int x, int z) {
  L9: assume z == x + 5;
      return; 
}

\end{lstlisting}}
    \end{minipage}
  \caption{An Example of a Passified Program}
  \label{Fi:ExampleProgram}
\end{figure}

\subsection{VC generation for a single procedure}
\label{Se:SingleProcVC}

Let $p(\vec{x})$ be a procedure that takes a sequence of arguments $\vec{x}$.
Further, assume that $p$ does not include procedure calls. In that case, we construct a
formula $VC(p)(\vec{x})$ such that $p$ has a terminating execution starting
from  arguments $\vec{c}$ if and only if $VC(p)(\vec{c})$ is satisfiable.

The VC is constructed as follows. For each block labelled $l$, let $b_l$ be a
fresh Boolean variable and $i_l$ be a unique integer constant. Let $\bsucc(l)$
be the set of successor blocks of $l$ (mentioned in the \textbf{goto} statement at the end
of block $l$, if any). Further, let $e_l$ be a conjunction of all assumed
expressions in the block. Let $\varphi_l$ be $(b_l \Rightarrow e_l)$ if
the block $l$ ends in a return statement, otherwise let it be:
\begin{equation} \label{Eqn:VCphi}
b_l \Rightarrow (e_l \wedge \bigvee_{s \in \bsucc(l)} (b_s \wedge (i_s
== f(i_l))))
\end{equation}
where $f$ is an uninterpreted function $\mathbb{Z} \rightarrow \mathbb{Z}$
called the control-flow function.

The variables $b_l$ are collectively referred to as \textit{control variables}.
Intuitively, $b_l$ is \textit{true} when control
reaches the beginning of block $l$ during the procedure's execution. The constraint
$\varphi_l$ means that if the control reaches block $l$, then it must satisfy
the assumed constraints on the block ($e_l$) and pick at least one successor
block to jump to. The function $f$ is simply to record chosen successor.

Let $l_p$ be the label of the first block of $p$ (where procedure execution begins). Let
$\blocks(p)$ be the set of block labels in $p$.  Then,
$VC(p)$ is $b_{l_p} \wedge \bigwedge_{l \in \blocks(p)} \varphi_l$.
If the VC is satisfiable, then one can read-off the counterexample trace from a satisfying
assignment by simply looking at the model for $f$.
As an example, the VC of procedure \texttt{foo} of \figref{ExampleProgram} is given in Figure~\ref{fig:VCfoo}.

The arguments of a procedure are its \textit{interface} variables and we make
these explicit in the VC. For instance, we will write $VC(\texttt{foo})(x, z)$
to make it explicit that $x$ and $z$ are the interface variables (free
variables) and the rest of the variables are implicitly existentially quantified.

\subsection{Stratified Inlining}
\label{Se:StratifiedInlining}

Inlining all procedure calls can result in an exponential blowup in program size. For
that reason, the \textit{stratified inlining} (SI) algorithm \cite{corral}
constructs the VC of a program in a lazy fashion. For ease in description,
assume that each block can have at most one procedure call.
For a procedure $p$, let $\pVC(p)$, called the
\textit{partial VC}, be the VC of the procedure constructed as described in the
previous section where each procedure call is replaced with an ``\textbf{assume}
\textit{true}'' statement.


\begin{figure}[t]
    \scriptsize
    \begin{minipage}{1\textwidth}
\begin{align*}
    VC(\texttt{foo}):~~~ & b_{L5} \\
  \land~~~    & b_{L5} \implies (b_{L6} \land f(5) == 6) \lor (b_{L7} \land f(5) == 7) \\
  \land~~~    & b_{L6} \implies d \land z == x + 1 \land b_{L8} \land f(6)  == 8 \\
  \land~~~    & b_{L7} \implies \lnot d \land z == x - 1 \land b_{L8} \land f(7)  == 8 \\
  \land~~~    & b_{L8} \implies true \\
\end{align*}
    \end{minipage}\\
    \caption{\label{fig:VCfoo} The VC of procedure \texttt{foo} from \figref{ExampleProgram}}
\end{figure}

Given that programs can only have assume statements, the partial VC of a
procedure represents an over-approximation of the procedure's behaviors, one
where it optimistically assumes that each callee simply returns. Similarly, for a procedure $p$,
if we replace each call with an ``\textbf{assume} \textit{false}'' statement,
then we get an under-approximation of $p$. The VC of this under-approximation
can be obtained by simply setting the control variables $b_l$ to false for each
block $l$ with an  ``\textbf{assume} \textit{false}'' statement. For instance,
$\pVC(\texttt{main})$ is an over-approximation of \texttt{main}, whereas the
following is an under-approximation:
$pVC(\texttt{main}) \land \lnot b_{L1} \land \lnot b_{L2} \land \lnot b_{L3}$.

\Omit{

\begin{align*}
VC(Main; foo) \;:\; & N5 \\
   \land            & N5 \implies N6 \lor N7 \\
   \land            & N6 \implies f \land (z == x + 1) \\
   \land            & N7 \implies \lnot f \land (z == x - 1) \\
VC(Main; bar) \;:\; & N9 \land N9 \implies (z == x + 5)\\
VC(Main; baz) \;:\; & N10 \land N10 \implies (y == 3)\\
VC(Main) \;:\;& N0 \\
   \land      & N0 \implies N1 \lor N2 \\
   \land      & N1 \implies c \land M_{foo} \land N3 \\
   \land      & N2 \implies \lnot c \land M_{bar} \land N3 \\
   \land      & N3 \implies M_{baz} \land N4 \\
   \land      & N4 \implies \lnot(z == 0)
\end{align*}
We finally connect all the procedural VCs and construct the VC of $P$ as:
\begin{align*}
VC(P)   \;:\; & VC(Main)\\
\land         & M_{foo} \implies VC(Main; foo) \\
\land         & M_{bar} \implies VC(Main; bar) \\
\land         & M_{baz} \implies VC(Main; baz) \\
\end{align*}
Now that we have explained VC generation, let us turn our attention towards a brief explanation of the Stratified Inlining (SI) algorithm.

}

A \textit{static callsite} is defined as the pair $(l, p)$
that represents the (unique) call of procedure $p$ in block $l$.
For instance, \texttt{main} of \figref{ExampleProgram} has three callsites:
$(\texttt{L1}, \texttt{foo})$, $(\texttt{L2}, \texttt{bar})$, $(\texttt{L3},
\texttt{baz})$. A \textit{dynamic callsite} is a stack of static callsites that
represents the runtime stack during a program's execution. We assume
that \texttt{main} is always present at the bottom of the stack for any dynamic
callsite. For instance, $[\texttt{main}, (\texttt{L1}, \texttt{foo})]$
represents the call stack where \texttt{main} executed to reach \texttt{L1} and
then called \texttt{foo}.

For a procedure $p$, let $\callsites(p)$ be the set of static callsites in $p$.
Given a static callsite $s$, and dynamic callsite $c$, let $s :: c$ be the
dynamic callsite where $s$ is pushed on the top of the stack $c$.
SI can require to inline the same procedure multiple times. Suppose that a
procedure $p$ calls $p'$ twice, once in block~$l_1$ and once in block $l_2$.
Dynamic callsites will help distinguish between the two instances of $p'$: the
first will have $(l_1, p')$ on top of the stack and the latter will have $(l_2,
p')$ on top of the stack.

We must take care to avoid variable name clashes between different VCs as we
inline procedures. For a dynamic callsite $c$ and procedure $p$ that is at the
top of $c$, let $\pVC(p, c)$ be the partial VC of $p$ (as described
earlier in the section), however for the construction of the partial VC, we
use globally fresh control variables (variables $b_l$ of
\eqnref{VCphi}), globally fresh block identifiers (constants $i_l$
of \eqnref{VCphi}) as well as globally fresh instances for the local
variables. In $\pVC(p, c)$, the argument $c$ is only used for
bookkeeping purposes: let $\cvar(l, c)$ refer to the control variable used for
block $l$ when constructing $\pVC(p, c)$. If $c$ is $(l,p) :: c'$, then let
$\cvar(c)$ be $\cvar(l', c')$. Similarly, if $p'$ is called from procedure $p$
in block $l'$, then let $\interfacevar((l', p') :: c)$ be the set of
interface variables (actuals) for the call to procedure $p'$ in block $l'$ in
$\pVC(p, c)$.

\Omit{
\begin{algorithm}[t]
\DontPrintSemicolon
\KwIn{A Program $P$ with starting procedure \texttt{main}}
\KwIn{An SMT solver $\solver$}
\KwOut{\textsc{Safe}, or \textsc{UnSafe}($\tau$)}
C $\gets$ $\{ [\texttt{main}, s] \mid s \in callsites(\texttt{main}) \}$ \label{Ln:OpenCallSitesInit} \\
$\solver$.Assert(pVC(\texttt{main}, [\texttt{main}])) \label{Ln:SolverInit} \\
  \While{true \label{Ln:Loop}}{
  \textit{// Under-approximate check}\\
  $\solver$.Push()\\
  \ForAll{$c \in C$} {
    $\solver$.Assert($\neg \cvar(c)$) \label{Ln:BlockCalls}\\
  }
  \If {$\solver$.Check() == \textsc{SAT} \label{Ln:UnderapproxCheck}} {
    \Return \textsc{UnSafe}($\solver$.Model())
  }
  $\solver$.Pop()\\
  \textit{// Over-approximate check}\\
  \If {$\solver$.Check() == \textsc{UNSAT} \label{Ln:OverapproxCheck}}{
    \Return \textsc{Safe}
  }
  \Else {
    $\tau \gets \solver$.Model() \label{Ln:Cex} \\
        $C' \gets C \cap \callsites(\tau)$ \label{Ln:ToExpand} \\
        $C'' \gets \emptyset$ \\
        \ForAll{$c \in C'$} {
          \textbf{let} $(l, p) :: c' = c$ \\
          $\solver$.Assert($\cvar(c) \implies \pVC(p,c)(\interfacevar(c))$) \label{Ln:Inline} \\
          $C'' \gets C'' \cup \{ s :: c \mid s \in \callsites(p) \}$ \label{Ln:NewCallsites} \\
        }
        $C \gets (C - C') \cup C''$ \label{Ln:UnionCallsites}\\
    }
  }
    \caption{The Stratified Inlining algorithm. \todo{remove this with Akash's concent}}
\label{Alg:SI}
\end{algorithm}
}

\begin{algorithm}[t]
\DontPrintSemicolon
\KwIn{A Program $P$ with starting procedure \texttt{main}}
\KwIn{An SMT solver $\solver$}
\KwOut{\textsc{Safe}, or \textsc{UnSafe}($\tau$)}
C $\gets$ $\{ [\texttt{main}, s] \mid s \in callsites(\texttt{main}) \}$ \label{Ln:OpenCallSitesInit} \\
$\solver$.Assert(pVC(\texttt{main}, [\texttt{main}])) \label{Ln:SolverInit} \\
  \While{true \label{Ln:Loop}}{
     $outcome \gets \textsc{SIStep}(P, C, \solver)$ \label{Ln:CallSIStep}\;
     \If {outcome == \textsc{Safe} $\vee$ outcome == \textsc{UnSafe}($\tau$)} {
            \Return $outcome$ \label{Ln:SIStepDecision}
      }
      \Else {
	  \textbf{let} \textsc{NoDecision}$(\_, \_, C') = outcome$ \\
          $C \gets C'$ \label{Ln:SIStepNoDecision}
      }
  }
\caption{The Stratified Inlining algorithm.}
\label{Alg:SI}
\end{algorithm}

\begin{algorithm}[ht]
\DontPrintSemicolon
\KwIn{A Program $P$, a set of callsites $C$}
\KwIn{An SMT solver $\solver$}
\KwOut{\textsc{Safe}, \textsc{UnSafe}($\tau$), \textsc{NoDecision}(uc, I, C)}
{
  \textit{// Under-approximate check}\\
  $\solver$.Push()\\
  \ForAll{$c \in C$} {
    $\solver$.Assert($\neg \cvar(c)$) \label{Ln:BlockCalls}\\
  }
  \eIf {$\solver$.Check() == \textsc{SAT} \label{Ln:UnderapproxCheck}} {
    \Return \textsc{UnSafe}($\solver$.Model())
}{
    $uc \gets \solver$.UnsatCore()\; \label{Ln:UnsatCoreCheck}
}
  $\solver$.Pop()\\
  \textit{// Over-approximate check}\\
  \If {$\solver$.Check() == \textsc{UNSAT} \label{Ln:OverapproxCheck}}{
      \Return \textsc{Safe}
  }
  \Else {
    $\tau \gets \solver$.Model() \label{Ln:Cex} \\
        $I \gets C \cap \callsites(\tau)$ \label{Ln:ToExpand} \\
        $C' \gets \emptyset$ \\
        \ForAll{$c \in I$} {
          $C' \gets \textsc{Inline}(c)$ \\
        }
        $C \gets (C - I) \cup C'$ \label{Ln:UnionCallsites}\\
        \Return \textsc{NoDecision}($uc$, $I$, $C$)
    }
  }
  \caption{\textsc{SIStep}($P$, $C$, $\solver$)}
\label{Alg:SIStep}
\end{algorithm}

\begin{algorithm}[t]
\DontPrintSemicolon
\KwIn{A dynamic callsite $c$, An SMT solver $\solver$}
\KwOut{A set of open callsites $C'$}
\textbf{let} $(l, p) :: c' = c$ \label{Ln:InlineBegin}\\
$\solver$.Assert($\cvar(c) \implies \pVC(p,c)(\interfacevar(c))$) \label{Ln:Inline} \\
$C' \gets C' \cup \{ s :: c \mid s \in \callsites(p) \}$ \label{Ln:NewCallsites} \label{Ln:InlineEnd} \\
\Return $C'$
\caption{\textsc{Inline}($c$, $\solver$)}
\label{Alg:inl}
\end{algorithm}

The SI algorithm is shown in \algoref{SI}. The algorithm requires an SMT solver with the usual interface. We use the
\textit{Push} API to set a backtracking point and a \textit{Pop} API that
backtracks by removing all asserted constraints until a matching \textit{Push}
call. Further, we assume that a counterexample trace can be extracted from a
model returned by the solver.

The algorithm works by
iteratively refining over-approximations of the program (in hope of getting an
early \textsc{Safe} verdict) and under-approximations of the program (in hope of
getting an early \textsc{UnSafe} verdict). Both these approximations are refined
by inlining procedures.

Line~\ref{Ln:OpenCallSitesInit} initializes a set $C$  of
\textit{open} dynamic callsites. This set represents procedure calls that have not
been inlined yet. The partial VC of \texttt{main} is asserted on the solver in
\lineref{SolverInit}. SI then iteratively calls \textsc{SIStep} (\algoref{SIStep})
that either returns a definitive verdict (\lineref{SIStepDecision})
or refines the set of open callsites (\lineref{SIStepNoDecision}).

The \textsc{SIStep} routine, shown in \algoref{SIStep}, does an under-approximate check
(\lineref{UnderapproxCheck}) by assuming that calls at each of the
open callsites cannot return (\lineref{BlockCalls}). If it finds a
counterexample trace, SI returns \textsc{UnSafe}. This trace is guaranteed to
only go through inlined procedure calls because all the open ones were blocked.
Ignore the call to gather the unsat core shown on \lineref{UnsatCoreCheck} for now; we use
this information in the next section.

Next, \textsc{SIStep} does an over-approximate check (\lineref{OverapproxCheck}). If this is
UNSAT, then SI returns \textsc{Safe}. If the check was satisfiable, then we
construct the counterexample trace from the model provided by the solver (\lineref{Cex}).
This trace is guaranteed to
go through at least one open call site (because the under-approximate check was
UNSAT). The SI algorithm proceeds to inline the procedures called at each of the open callsites
that the trace goes through. Such callsites are recorded in variable $I$
(\lineref{ToExpand}); these get returned for bookkeeping purposes (used in the
next section). Callsites in $I$ are inlined by asserting the partial VC of
the callee, as shown in \lineref{Inline} in \algoref{inl}. Read the asserted constraint as follows:
if the control variable of the calling block is set to \textit{true} then
the VC of the procedure must be satisfied. The use of $\interfacevar$
ensures that formals are substituted with actuals for the procedure call.
New callsites that are created as a
result of the inlining are recorded in $C'$ and then eventually added back to
$C$ (\lineref{UnionCallsites}). The control returns to SI and the process then repeats.
An example illustrating the execution of SI is shown in Table~\ref{Tab:SIExample}.


Define a \textit{call tree} to be a (prefix-closed) set of dynamic callsites that represents
all dynamic callsites that have been inlined by the SI algorithm at any point in time.
We call this set as a tree because it can be represented as an unfolding of the
program's call graph.

\begin{table}[t]
  \centering
  \scriptsize
  \begin{tabular}{l| l| l| r}
    \hline
    \textsc{SIStep} & Action & Open Callsites & Inlined Callsites \\
    \hline
    \multirow[t]{ 3}{*}{Step-0} & Assert $\pVC(\texttt{main})$ & [\texttt{main}, (\texttt{L1},\texttt{foo})], & [\texttt{main}]\\
							       &                  & [\texttt{main}, (\texttt{L2},\texttt{bar})], &     \\
				   &                  & [\texttt{main}, (\texttt{L3},\texttt{baz})]  &     \\
    \hline
    \multirow[t]{ 4}{*}{Step-1} & Underapprox check: \texttt{UNSAT}   & & \\
                                & Overapprox check: \texttt{SAT}      & & \\
                                & Assert $\pVC(\texttt{foo})$   & [\texttt{main}, (\texttt{L2},\texttt{bar})] & [\texttt{main}, (\texttt{L1},\texttt{foo})]\\
                                & Assert $\pVC(\texttt{baz})$   &                   & [\texttt{main}, (\texttt{L3},\texttt{baz})]\\
    \hline
    \multirow[t]{ 4}{*}{Step-3} & Underapprox check: \texttt{SAT}     & [\texttt{main}, (\texttt{L2},\texttt{bar})] & \\
                                & Return \textsc{Unsafe}              &                   & \\
    \hline
  \end{tabular}
  \caption{Execution of SI on the program of Fig. \ref{Fi:ExampleProgram}}\label{Tab:SIExample}
\end{table}

\section{Splitting the Search}
\label{Se:Algorithm}

\toolname{} employs a \textit{decomposition}-based strategy to achieve
parallelism. During the course of execution of the SI algorithm, \toolname{}
\textit{splits} the current verification task by picking a dynamic callsite $c$ that
has already been inlined by SI. This generates two \textit{partitions}:
one that requires executions to pass through $c$ (referred to as the
\textit{must-reach} partition),
and the other that requires executions to avoid $c$ (referred to
as the \textit{must-avoid} partition). This strategy provides for an exhaustive
and \textit{path-disjoint} partitioning of the search space.


Formally, a \textit{partition} is a pair
$(T, D)$ where $T$ is a call tree (i.e., set of inlined callsites) and $D$ is a
set of \textit{decisions} (either $\ma(c)$ or $\mr(c)$ for $c \in T$).
As a notation shorthand, for a partition $\rho = (T, D)$ and callsite $c$, let $\rho + c$
be the partition $(T \cup \{ c \}, D)$. Similarly, for a decision $d$, let $\rho
+ d$ be the partition $(T, D \cup \{ d \})$. Further, let $\calltree(\rho) = T$
and $\decisions(\rho) = D$.
One can also see the above strategy as dividing the proof obligation
(correctness theorem) on the complete program into a set of \textit{lemmas}
corresponding to each of the partitions. 

This section addresses two primary concerns: $(a)$
how to enforce splitting decisions during search? (\sectref{Constraints}),
and $(b)$ how to choose a callsite for splitting? (\sectref{SplittingHeuristic}).

\subsection{Encoding splitting decisions in SI as constraints}
\label{Se:Constraints}

The constraint for $\ma(c)$ is relatively straightforward. It is simply $\neg \cvar(c)$.
Asserting this constraint any time after SI has inlined $c$ will ensure
that control cannot go through $c$, thus SI will avoid $c$ altogether.

We next describe the encoding of the $\mr$ constraint by first looking at the
single-procedure case. For a procedure
$p$, we introduce \textit{must-reach} control variables $r_l$, one for each
basic block $l$ of $p$.
Intuitively, setting $r_l$ to \textit{true} should mean that control must go
through block $l$. Recall from \sectref{Background} that the VC of a procedure uses
$i_l$ as a unique integer constant for block $l$ and $f$ as the control-flow
function. We define $\mr(p)$ as the following constraint:

\begin{equation} \label{Eqn:MustReach}
  \bigwedge_{l \in \blocks(p)} (r_l \Rightarrow \bigvee_{n \in \bpred(l)} (r_n
  \wedge f(i_l) == i_n))
\end{equation}

This constraint enforces that if a block $l$ must be reached, then one of its
predecessors must be reached. The use of the control-flow function ties this
constraint with the procedure's VC.
For any block $l$, asserting $r_l \wedge \mr(p)$, in addition to the VC of $p$
will enforce the constraint that control \textit{must} pass through block $l$.
The proof is straightforward and we omit it from this paper.

For multi-procedure programs, we construct the $\mr$ constraint inductively.
Let $\mr(p, c)$ be the constraint $\mr(p)$, but where the block identifiers
$\{i_l\}$ are the same as the ones used in $\pVC(p, c)$. We construct $\mr(c)$
inductively over the length of $c$. If $c = [\texttt{main}]$, then
$\mr(c)$ is \textit{true}. Otherwise, if $c
= (l,p) :: c'$, then $\mr(c)$ is $r_l \wedge \mr(p, c') \wedge \mr(c')$.

\subsection{Choosing a splitting candidate}
\label{Se:SplittingHeuristic}


Given an unsatisfiable formula $\Phi$, expressed as a conjunction set of clauses $\{\phi_i\}$,
a \textit{minimal unsatisfiable core} (\textit{min-unsatcore}) is a subset of clauses $\Psi \subseteq \Phi$ whose conjunction is
still unsatisfiable and every proper subset of $\Psi$ is satisfiable. 

Consider the under-approximate check made by SI (\lineref{UnderapproxCheck} of
\algoref{SIStep}) where it blocks open-callsites and attempts to find a
counterexample in the currently inlined portion of the program. This check is 
a conjunction of constraints, passed via $\solver.\texttt{Assert}$, of two
forms. First is the (partial) VCs of inlined callsites (\lineref{Inline} of \algoref{inl})
and second is the blocked open callsites (\lineref{BlockCalls} of \algoref{SIStep}).
If the check is unsatisfiable, then we extract its \textit{min-unsatcore} and
represent it as a set of callsites $uc$ (that may be inlined or may be open).
The set $uc$ represents the current proof of
safety of the program. Inlined callsites that are not part of $uc$ are deemed
\textit{search-irrelevant} because whether they were inlined or not is immaterial to
conclude safety of the program (at this point in the search). Formally, those
callsites could have been left open (i.e., over-approximated) and the check would still
be unsatisfiable. Therefore, the solver is likely to spend its energy searching
and expanding the $uc$ portion of the calltree as the search proceeds further. Consequently, 
we restrict splitting to a callsite chosen from $uc$ so that we split where the
search complexity lies.



\begin{figure}[t]
  \includegraphics[scale=0.7]{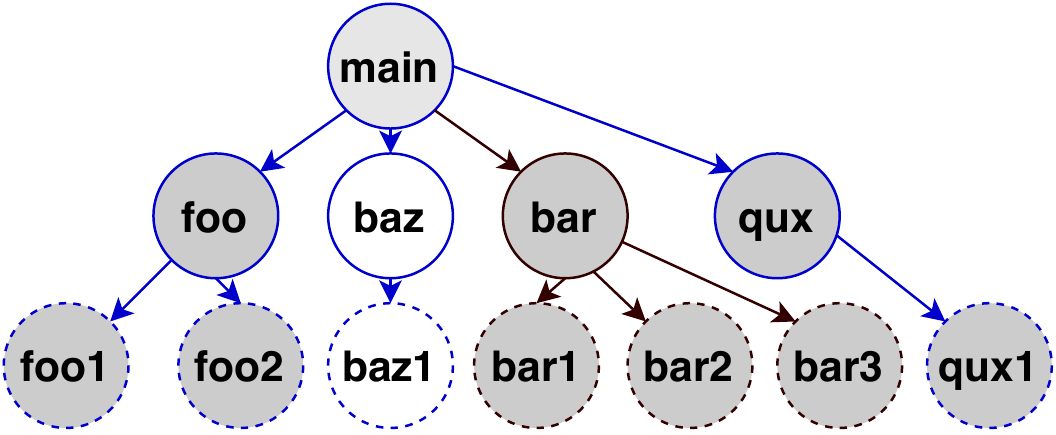}
    \caption{\label{fig:splitting}Proof-guided splitting}\label{split}
\end{figure}

Consider the inlining tree shown in Figure~\ref{fig:splitting}, where the open callsites appear as dotted circles and the
inlined ones are shown as solid circles; the shaded nodes are the callsites that
appear on the \textit{min-unsatcore} ($uc$).
\Omit{
Among the open callsites, \texttt{baz1} does not appear on the
\textit{min-unsatcore}; hence, by the property of \textit{min-unsatcore} mentioned
above, overapproximating \texttt{baz1} will still maintain the query as
unsatisfiable, indicating that \texttt{baz1} may not be relavant for the
property under consideration. Intuitively, the callsites in the
\textit{min-unsatcore} capture parts of the program that induce behaviours
affecting the property of interest, and hence, are the current focus of the
stratified inlining search.
We, therefore, refer to the set of callsites appearing in the
\textit{min-unsatcore} from an unsatisfiable under-approximate query as
\textit{search-relevant} callsites. We use search-relevancy as a metric to
identify split-candidates.
}
In this case, both \texttt{baz} and \texttt{baz1} are ruled out for falling
outside $uc$. If we pick some other callsite to split, say \texttt{qux}, then
the $\mr(\texttt{qux})$ partition of that split is likely to search in the subtree rooted at
\texttt{qux}, whereas the $\ma(\texttt{qux})$ partition will search the $uc$ portion excluding
the subtree rooted at \texttt{qux}. We use a simple heuristic that roughly
balances these partitions. Let the current inlined calltree be $T$ and let $\texttt{subtree}(T, c)$ be the
subtree rooted at $c$. We choose the splitting callsite as the one that has
maximum number of relevant callsites in its subtree (excluding \texttt{main}
because that would be a trivial split). Formally, the splitting callsite is:
$$\underset{c \in uc}{\texttt{argmax}} \{ | \texttt{subtree}(T, c)\cap uc| \}$$
In our example, we will pick \texttt{bar} for splitting.

We note that this choice of balancing the partitions is just a heuristic. In
general, there may be dependencies between callsites. For instance, blocking one
callsite can block others or make others be \textit{must-reach} because of control-flow
dependencies in the program. Our heuristic does not capture these dependencies. Furthermore, in
our implementation, we do not insist on obtaining a \textit{minimal} unsat core
in order to reduce the time spent in computing it.
Solvers generally provide a best-effort unsat core minimization (e.g.,
the \texttt{core.minimize} option in Z3).

\section{\toolname{} Design and Implementation}
\label{Se:DesignImplementation}

\toolname{} employs a client-server distributed architecture with a single
server and multiple clients. The server (\sectref{Server}) is responsible for coordination while
verification happens on the clients (\sectref{Client}). A client can decide to split its current
search, at which point it sends one partition to the server while it continues
on the other partition. If a client finishes its current search with a
\textsc{Safe} verdict, it contacts the server to borrow a new partition that it
starts solving.

\subsection{Client Design}
\label{Se:Client}

\begin{algorithm}[t]
\DontPrintSemicolon
\KwIn{A Program $P$}
\KwIn{An SMT solver $\solver$}
\While{true \label{Ln:ClientInfiniteLoop}}  {
    $\rho \gets \callServer$(\texttt{GET\_PARTITION}) \; \label{Ln:GetPartition}
    outcome $\gets$ \textsc{Verify}($P$, $\rho$, $\solver$) \; \label{Ln:CallVerify}
    $\callServerA$(\texttt{OUTCOME}, outcome) \; \label{Ln:ReturnOutcome}
    }
\caption{Client-side verification algorithm}
\label{Alg:Client}
\end{algorithm}

All clients implement \algoref{Client}. We use $\callServer$ as a
message-response interaction with the server. $\callServerA$ is the asynchronous
version where a message is sent to the server but a response is not expected.
A client repeatedly requests the server for a partition (\lineref{GetPartition}),
solves it (\lineref{CallVerify}) and sends the result back to the server on
completion. Each client uses its own dedicated SMT solver ($\solver$) for verification.

\textsc{Verify}~(\algoref{Verify}) maintains a stack of decisions $dstack$ and
a set of open callsites $C$. It starts off by preparing the input partition
(\linesref{SetupBegin}{SetupEnd}): it inlines the calltree of $\rho$ and asserts
all its splitting decisions.
The client then enters a verification loop (\lineref{VerifyLoop} that repeatedly
uses \textsc{SIStep} (\lineref{VerifyCallSIStep}) to expand its search.
If a counterexample is found (\lineref{VerifyUnsafe}), the client returns an \textsc{UnSafe} verdict back to
the server. If \textsc{SIStep} returns
\textsc{NoDecision}, it implies that some more procedures were inlined but the
search remained inconclusive; in this case, we perform the necessary bookkeeping
on the set of currently open callsites ($C'$), new procedures inlined ($I$),
and the minunsatcore from the unsat query ($uc'$).

If \textsc{SIStep} returned \textsc{Safe}, then the search on the current partition has finished and
the client must pick another partition to solve. This is done by returning the
\textsc{Safe} verdict (\lineref{VerifySafe}). The check on
\lineref{PartitionSteal} is an optimization that we describe later in this
section.

After checking the outcome of \textsc{SIStep}, the client decides if it is time
to split its search. This is referred abstractly as ``\textit{TimeToSplit}'' on
\lineref{whensplit}: the exact time is communicated by the server to client (see
\sectref{SplitRate}). For splitting, the client picks a callsite $c$
in accordance with our proof-guided splitting heuristic (from \sectref{SplittingHeuristic})
using the stored unsatcore $uc$. We note that the correctness of our technique
does not rely on when a split happens or what splitting callsite is chosen.
Therefore, these decisions can be guided by heuristics and tuned to optimized
performance.

\begin{algorithm}[t]
\small
\DontPrintSemicolon
\KwIn{A Program $P$, A partition $\rho$ of $P$, A solver $\solver$}
\KwOut{\textsc{Safe}, or \textsc{UnSafe}($\tau$)}
$\solver$.reset(),  $dstack \gets [\ ]$, $C \gets \emptyset$\, $uc \gets \emptyset$\;\label{Ln:SolverReset}
// \textit{Setup input partition} \;
\ForAll{$c \in \calltree(\rho)$ \label{Ln:SetupBegin}}{
  $C' \gets \textsc{Inline}(c)$, $C \gets (C - \{ c \}) \cup C'$\;
}
\ForAll{$d \in \decisions(\rho)$}{
	\textbf{if } $d == \texttt{MUSTAVOID}(c)$ \textbf{then } $\solver$.Assert($\ma(c)$)\;
	\textbf{if } $d == \texttt{MUSTREACH}(c)$ \textbf{then } $\solver$.Assert($\mr(c)$)\; \label{Ln:SetupEnd}
  }
\While{true \label{Ln:VerifyLoop}} {
	   outcome $\gets$ \textsc{SIStep}($P$, $C$, $\solver$)\; \label{Ln:VerifyCallSIStep}
      \If{outcome == \textsc{UnSafe}($\tau$) \label{Ln:VerifyUnsafe}} {
            \Return outcome\;
        }
        \ElseIf {outcome == \textsc{NoDecision}$(uc', I, C')$} {
            $uc \gets uc'$, $C \gets C'$, $\rho \gets \rho + I$\;
        }
        \Else { \label{Ln:Safe}
		\If{$\callServer$(\texttt{POP})==\texttt{YES} \label{Ln:PartitionSteal}} {
			\Repeat{$d$ == \texttt{MUSTAVOID} \label{Ln:PopLoop}}{
                    \textbf{let} $d(c) :: ds = dstack$\;
                    $\solver$.Pop(), $dstack \gets ds$, $\rho \gets \rho - d(c)$\;
                   }
                $\solver$.Push(), $\solver$.Assert(\mr(c)), $dstack \gets \texttt{MUSTREACH}(c) :: dstack$, $\rho \gets \rho + \texttt{MUSTREACH}(c)$\; \label{Ln:StealEnd}
            }
            \Else {
		    \Return outcome\; \label{Ln:VerifySafe}
            }
        }
        \If{\textit{TimeToSplit}} { \label{Ln:whensplit}
      $c \gets \textbf{choose}(\calltree(\rho), uc)$\; \label{Ln:findsplit}
      $\solver$.Push()\; \label{Ln:SplitBegin}
	    $\solver$.Assert($\ma(c)$)\;
            $\callServerA$(\texttt{SEND\_PARTITION}, $\rho + \texttt{MUSTREACH}(c)$)\;\label{Ln:SplitSend}
            $dstack \gets \texttt{MUSTAVOID}(c) :: dstack$, $\rho \gets \rho + \texttt{MUSTAVOID}(c)$\; \label{Ln:SplitEnd}
        }
    }
\caption{\textsc{Verify}($P$, $\rho$, $\solver$)}
\label{Alg:Verify}
\end{algorithm}



After splitting, the client continues along the partition with the
$\texttt{MUSTAVOID}(c)$ decision (let's call this partition $\rho_1$). The other
partition ($\rho_2$) is sent to the server
(\lineref{SplitSend}). Note further that on \lineref{SplitBegin}, the client
creates a backtracking point that is \textit{just before} the decision on $c$ is
asserted. This backtracking point is exploited in
\linesref{PartitionSteal}{StealEnd}. When the client finishes search on
$\rho_1$, it pings the server to know if $\rho_2$ has already been solved by a
different client or not. If not, it simply backtracks the solver state and
asserts the flipped decision $\texttt{MUSTREACH}(c)$ to immediately get set up
for search on $\rho_2$. This way, the client avoids the expensive setup of
initializing a new partition (\linesref{SetupBegin}{SetupEnd}). Because
splitting can happen multiple times, the loop on \lineref{PopLoop}
is necessary to follow along the recorded stack of decisions.

\subsection{Server Design}
\label{Se:Server}

We assume that each client has an associated unique identifier. Each message coming from a
client is automatically tagged with the client's identifier. The server
maintains two data structures. The first is an array $Q$ of double-ended queues.
The queue $Q[id]$ stores all partitions produced by client $id$. The second is a
queue $wt$ of clients that are currently idle.

\begin{figure}[t]
  \centering
  \includegraphics[scale=0.6]{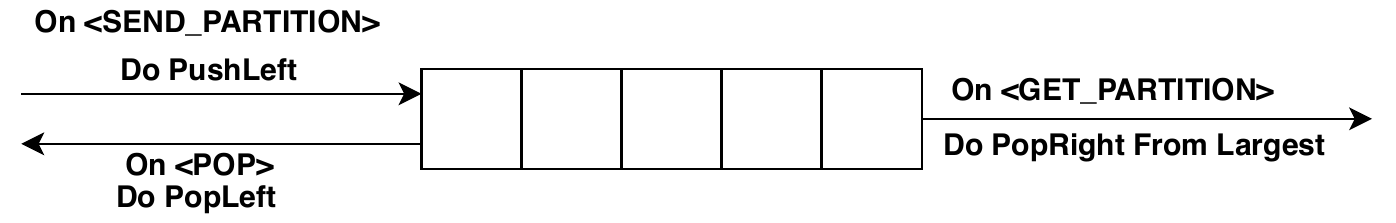}
  \caption{\label{Fi:deqFig}Maintaining the double-ended queues}
\end{figure}

The server processes incoming messages as follows. On receiving the message
$\langle\texttt{SEND\_PARTITION}, \rho\rangle$ from client $id$, it does a
\texttt{push-left} to insert $\rho$ into $Q[id]$. (The manipulation of $Q$ is
depicted in \figref{deqFig}.) This ensures that latest partitions (which have
a larger number of decisions and a larger call tree) from a particular client $id$ appear on the
left of $Q[id]$.

On receiving message $\langle\texttt{GET\_PARTITION}\rangle$
from client $id$, the server needs to reply with a partition because $id$ has
just become idle. If all queues $Q[i]$ are empty, then $id$ is inserted into
$wt$ and the client is kept waiting for a reply. Otherwise, the server picks the
longest queue $Q[i]$, does a \texttt{pop-right} and replies to the client. This
strategy attempts to avoid skew in queue sizes. Further, the rightmost partition
is the smallest in that queue, which mimizes the setup time for that partition
for the client that will get it. As more
partitions are reported to the server (via a $\texttt{SEND\_PARTITION}$),
the server loops through $wt$, replying to as many idle clients as possible with
partitions popped-right from the currently longest queue.

The message $\langle\texttt{POP}\rangle$ from client $id$ implies that the client
wishes to backtrack to its previously reported partition. Because reported
partitions are pushed-left, and other clients (on $\texttt{GET\_PARTITION}$)
steal from the right, the previously reported partition from client $id$ is exactly
the leftmost one in $Q[id]$, if any. Thus, the server replies \texttt{YES}
back to the client if $Q[id]$ is non-empty, followed by a \texttt{pop-left}. Otherwise, the
server replies \texttt{NO}.

The server additionally listens to \texttt{OUTCOME} messages. If any client
reports \textsc{Unsafe}, all clients are terminated and the \texttt{UnSafe}
verdict is returned to the user. The server returns \textsc{Safe} verdict to the
user when all queues in $Q$ are empty and all clients are idle (i.e., $wt$
consists of all clients).

Our design of the work-queue $Q$, as an array of sorted (by size) work-queues, is in contrast
with using a centralized queue that is standard in classical work-stealing
algorithms. It is useful for avoiding skew in queue sizes, distributing smaller
partitions first, and enabling the client-backtracking optimization.

%

\subsection{Adaptive rate of splitting}
\label{Se:SplitRate}

While a low splitting rate inhibits parallelism, a high rate increases the
partition-initialization overhead on the clients. \toolname{} uses a dynamic
split-rate determined by the number of idle clients and the number of partitions available at the server.
Each client maintains a \textit{split time interval} $\delta$ (in seconds) and
splits the search (``\textit{TimeToSplit}'' of \algoref{Verify}), if $\delta$ seconds have elapsed since the last split.
The value of $\delta$ starts as a constant $\delta_c$ and is
updated by the server as follows:

\begin{equation}\label{sprate}
\delta_i =
    \begin{cases}
      \frac{Q[i].count}{wt.count}\times \delta_c & \mbox{if } wt.count \neq 0 \\
      K \times \delta_c, & \mbox{otherwise}.
    \end{cases}
\end{equation}

In the first case, a client's splitting is slowed down in proportion to its queue size
(divided by the number of idle clients).
The second case applies when there are no idle clients. Increasing $\delta$ by a
factor of $K$ reduces the rate of splitting drastically. We use
$\delta_c=0.5s$ and $K=20$ in our experiments.


\section{Experimental Results}
\label{Se:Experiments}

%

We evaluated \toolname{} on SDV benchmarks~\cite{sdvbench}, compiled from
real-world code that exercises all features of the C language: loops and recursion (up to a bounded depth), 
pointers, arrays, heap, bit-vector
operations, etc. The performance of \toolname{} was compared against \corral{}
\cite{corral} that implements the sequential Stratified Inlining algorithm.
\corral forms a good baseline because it has been optimized heavily for SDV over
the years \cite{DBLP:conf/sigsoft/LalQ14}.

We only selected hard benchmarks (where \corral took at least 200 seconds to solve
or timed out). We ran \toolname{} with $32$ clients. Timeout was set to $1$ hour. 
We conducted our experiments with the server running on one
machine (16 core, 64 GB RAM) and the 32 clients running on another machine
(72-core with Intel Xeon Platinum 8168 CPU and 144 GB RAM), communicating via
HTTP calls. As clients never communicate amongst themselves, this setup is equivalent
to running clients on different machines.



\begin{figure*}[t]
  \centering
  \subfloat[Scatter plot of running
  times]{\label{Fi:timeScatter}\includegraphics[width=0.31\textwidth]{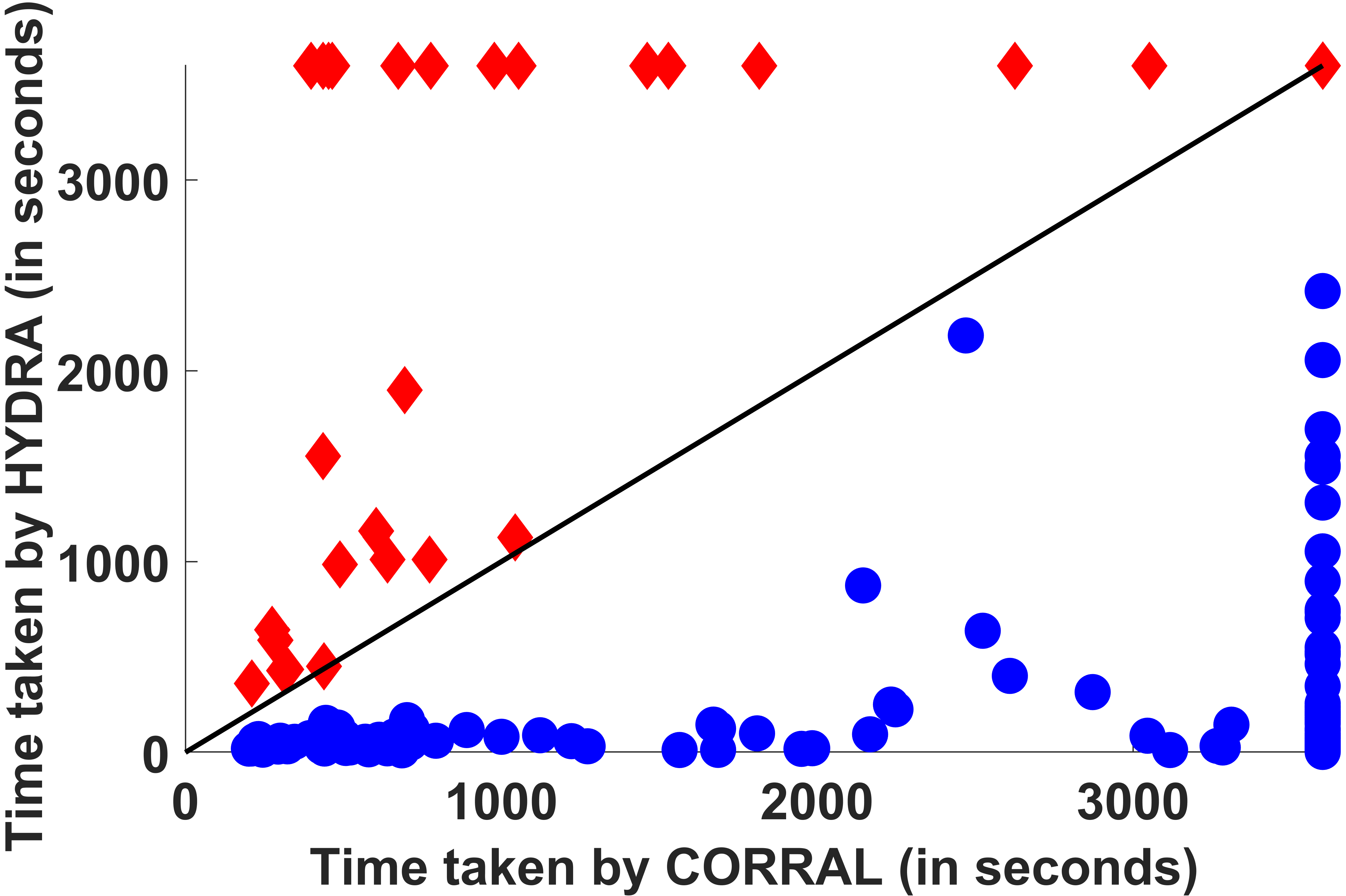}}\quad
  \subfloat[Histogram of speedup of $\toolname$ over
  $\corral$]{\label{Fi:barSpeedup}\includegraphics[width=0.31\textwidth]{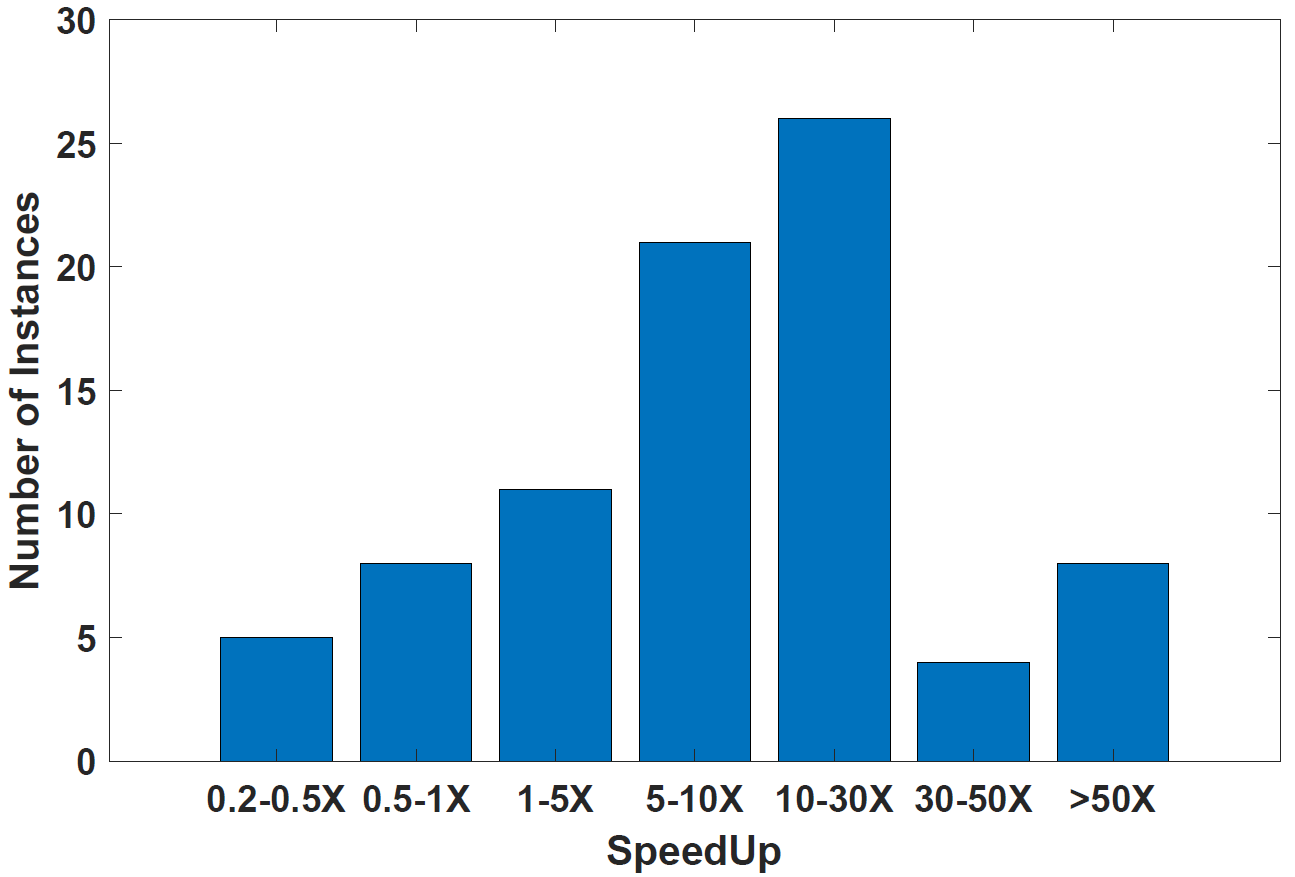}}\quad
  \subfloat[Cactus plot of instances
  solved]{\label{Fi:corralCactus}\includegraphics[width=0.31\textwidth]{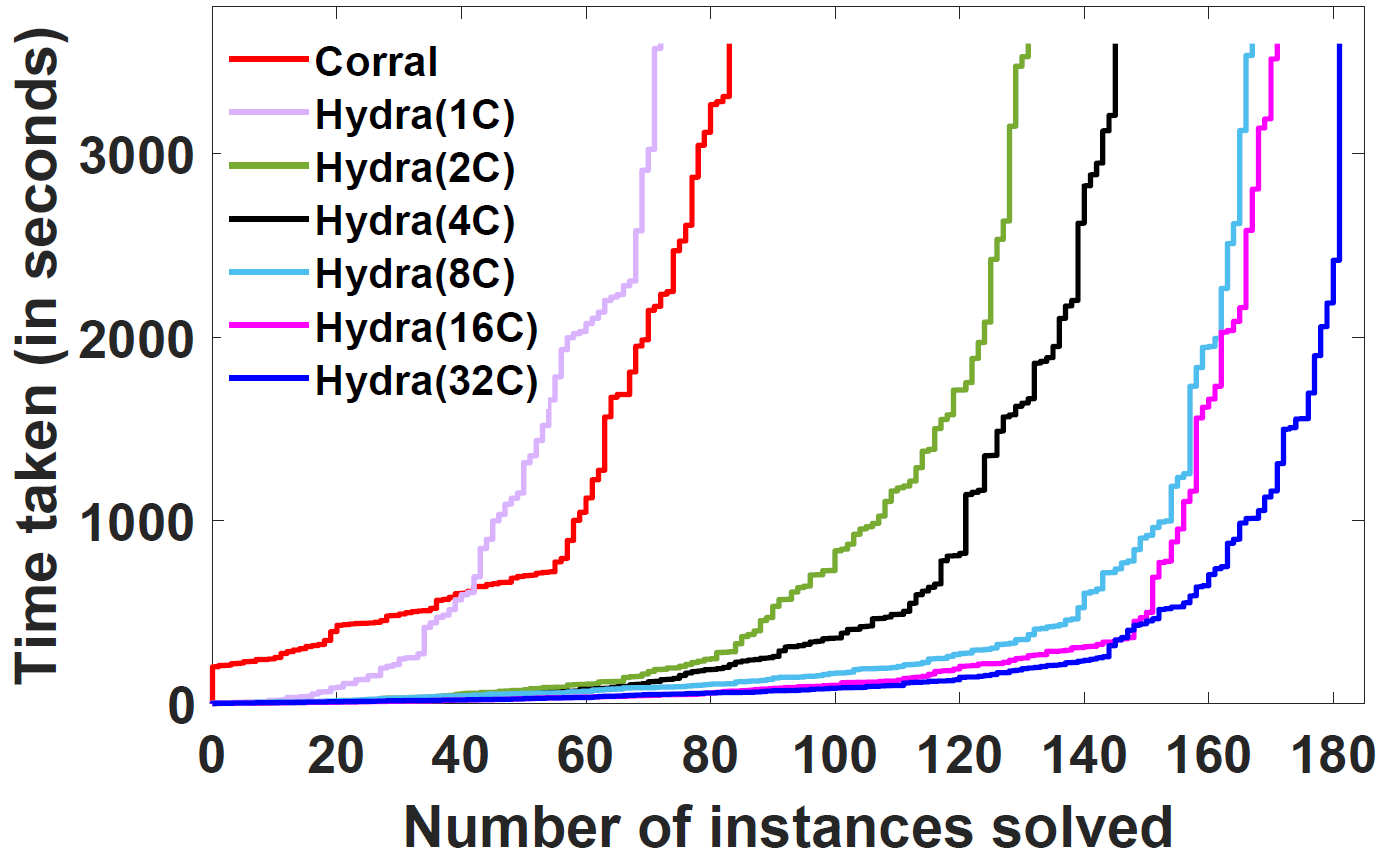}}
    \caption{Comparison of {\toolname} against \corral on SDV benchmarks}
\end{figure*}

\subsection{\toolname{} versus \corral{}}
\noindent \textit{Instances Solved.~}
There were a total of $333$ programs. \toolname{} solved 99 instances (30\%) on
which \corral timed out (34 of these were \textsc{Safe} and the rest 65 were
\textsc{Unsafe}). Conversely, \corral solved 12 (4\%) instances on which \toolname{} timed
out. We did not investigate these cases in detail; in a practical scenario one
can simply dedicate a single client to run $\corral$ and get the best of both
tools. Overall, \toolname{} solved 183 (55\%) instances while \corral{} solved only 96 (29\%) instances.
Interestingly, there were 138 instances (41\%) that were unsolved by both \toolname{} and
\corral indicating the need for further improvements.





\noindent \textit{Verification Time.~}
In terms of running time, \toolname{} was significantly faster than \corral
in most (84\%) cases: \figref{timeScatter} shows the scatter plot of running
times. \figref{barSpeedup} is a histogram of the speedup of \toolname{} over
\corral{}.  For
example, there were $8$ instances where \toolname{} was more than $50\times$
faster than $\corral$. A small fraction of instances had slowdowns as well, but
the worst among these was $0.2\times$, i.e., \corral was $5\times$ faster than
\toolname{}. Over all instances, the mean speedup is $20.4\times$ and median
speedup is $9.7\times$.
Speedup excludes cases in which one of the tools timed out. 

\noindent \textit{Scalability.~}
\figref{corralCactus} is a cactus plot illustrating the scalability of
\toolname{} with the number of clients. \corral{} is able to solve only 58 instances
within 1000 seconds. As expected, running \toolname{} with only a single client
results in worse performance than \corral (solves only 46 instances within 1000 seconds). 
However, the performance
improves significantly with the number of clients (solves 166 instances with 32
clients within 1000 seconds).


\subsection{Effectiveness of proof-guided splitting}

\noindent \textit{Empirical Analysis.~}
We define \textit{dissimilarity} $\eta(i, j)$ of a client $i$ with respect to client $j$ as 
$1 - \frac{|\mathcal{L}_i \cap \mathcal{L}_j|}{|\mathcal{L}_i|}$, where
$\mathcal{L}_i$, $\mathcal{L}_j$ denote the set of callsites that $i$ and $j$ have
inlined, respectively, when $\toolname$ finishes. A high value of $\eta(i, j)$ implies that the clients did
a different search. Note, however, that
$\eta(i, j)$ will never be $1$ because certain callsites (like \texttt{main}) 
will always need to be inlined by each client.  

Across all benchmarks and all client pairs, the average dissimilarity value was
0.55. This indicates enough difference among the inlined calltrees across
clients. 

\noindent \textit{Statistical Analysis.~}
We  implemented a randomized splitting algorithm that (1) decides to split/not-to-split at each inlining step uniformly at random, (2) if it has decided to split, it selects the splitting call-site uniformly at random.

We ran this randomized splitting algorithm 5 times for each program and compared the minimum verification time of these~5 runs for each instance against that of \toolname{}. Using the Wilcoxon Sign Rank test, we found that \toolname{} is statistically better than the randomized splitting algorithm with a p-value of 0.0012, indicating that the performance of the splitting heuristic is not accidental.

\subsection{Server optimizations}

We measured the performance impact of the server-side queue implementation on
\toolname{}. We compared our double-ended queues $Q$ from \sectref{Server}
against a classical work-stealing queue implementation. Our
implementation allowed \toolname{} to complete on 40\% more cases where using
the classical version made $\toolname$ time out. Further, \toolname's performance was 
8.5 times faster when both implementations terminated with a
verdict.

In terms of controlling the splitting rate, both the performance 
(p-value of $5.27\times 10^{-5}$) and the number of splits 
(p-value of $5.43\times 10^{-33}$) were found to be statistically better with
split-rate feedback.

\section{Related Work}
\label{Se:RelatedWork}

\noindent \textit{Parallelizing SAT/SMT solvers.~}
In contrast to parallelizing verification tasks, parallelizing SAT/SMT solvers has attracted wider attention. There have been two popular, 
incomparable~\cite{Marescotti2018}, approaches to parallelizing satisfiability problems:
portfolio-based techniques~\cite{Chaki2016,Hyvarinen2008,Wintersteiger2009} and
divide and conquer techniques (decomposition~\cite{Niklas2004,Hamadi2011} or
partitioning~\cite{Zhang1996,Martins2010,Bohm1996,Jurkowiak2001}). 
Portfolio-based strategies either run multiple different algorithms or multiple instances of a
randomized algorithm. They tend to work well in the presence of heavy-tailed
distribution of problem hardness.


Divide and conquer strategies are most similar to our work. They 
either use static partitioning, based on the structure of 
the problem~\cite{Marescotti2017}, or dynamic partitioning~\cite{Martins2010} 
based on run-time heuristics. However, unlike partitioning on individual
variables at the logical-level, we split at the program-level based on its
call graph. In our setting, the VC of a program can be exponential in the size
of the program. This makes it hard to directly use parallelized solvers; we 
must split even before the entire VC is generated. Furthermore, parallelized 
solvers are still not as mainstream as sequential solvers. Using solvers as 
a black-box allows us to directly leverage continued improvements in solver technology

\Omit{
In spirit, our algorithm is similar to many of the above SAT/SMT solvers employing the partitioning
strategy, especially the ones using guiding-paths~\cite{Schubert2005,Zhang1996}, using dynamic task
partitioning using sophisticated heuristics (like use of VDIS to select partitioning
variables~\cite{Martins2010,Plaza2006}) and work stealing (similar to proposals
like~\cite{Schubert2005} where long guiding paths are preferred for assigning work to an idle
client).
}

\Omit{
    However, as opposed to satisfiability, we target the problem of software verification,
demonstrating results for end-to-end program verification tasks. We use the solver as a black-box,
creating the splits at the level of the program's call graph, and as such can directly leverage any
advancements in parallel SMT solvers.
Furthermore, we use a decentralized work-queue (unlike most of the previous approaches) where the clients generate new
partitions and the server only orchestrates the queued-up tasks, while maintaining
per-client queues instead of a central task-queue.}

\noindent \textit{Parallelizing program verification.~}
Saturn \cite{aiken2007overview} is one of the earlier attempts at parallelizing
program verification. Saturn performs a bottom-up analysis on the call graph, 
generating summaries of procedures in parallel. While the intra-procedural analysis
of Saturn is precise, it only retains \textit{abstractions} of function
summaries, thus cannot produce precise refutations of assertions like BMC.

There have been attempts at parallelizing a top-down abstraction-based verifier 
\cite{albarghouthi2012parallelizing}  as well as the property-directed reachability
(PDR) algorithm~\cite{Bradley2011,Een2011,Marescotti2017,Chaki2016} and k-induction~\cite{Kahsai2011, Blicha2020}.
These all rely on the discovery of inductive invariants for proof generation, a
fundamentally different problem than BMC. It would be interesting future work 
to study the relative speedups obtained for parallelization in these respective
domains.

Closer to BMC, parallelization has been proposed by a partitioning of the
control-flow graph \cite{ganai2008d}. This approach does static
partitioning (based on program slicing) and does not consider procedures
at all (hence, must rely on inlining all procedures). Further, it has only been evaluated
on a single benchmark program. Our technique, on the other hand, performs
dynamic partitioning, supports procedures and has been much more extensively
evaluated.

\bibliographystyle{IEEETran}
\bibliography{references}
\end{document}